\documentclass[aip, pof,graphicx]{revtex4-2}

\usepackage{graphicx}
\usepackage{dcolumn}
\usepackage{bm}
\usepackage[utf8]{inputenc}
\usepackage[T1]{fontenc}
\usepackage{mathptmx}
\usepackage{color}
\usepackage{longtable}
\draft 

\begin{document}


\title[Force distribution within a barchan dune]{Force distribution within a barchan dune\\
This article may be downloaded for personal use only. Any other use requires prior permission of the author and AIP Publishing. This article appeared in Phys. Fluids 33, 013313 (2021) and may be found at https://doi.org/10.1063/5.0033964} 



\author{Carlos A. Alvarez}
\author{Erick M. Franklin}%
 \email{erick.franklin@unicamp.br}
 \thanks{Corresponding author}
\affiliation{ 
School of Mechanical Engineering, UNICAMP - University of Campinas,\\
Rua Mendeleyev, 200, Campinas, SP, Brazil
}%


\date{\today}

\begin{abstract}
Barchan dunes, or simply barchans, are crescent-shaped dunes found in diverse environments such as the bottom of rivers, Earth's deserts and the surface of Mars. In a recent paper [Phys. Rev. E 101, 012905 (2020)], we investigated the evolution of subaqueous barchans by using computational fluid dynamics - discrete element method (CFD-DEM), and our simulations captured well the evolution of an initial pile toward a barchan dune in both the bedform and grain scales. The numerical method having shown to be adequate, we obtain now the forces acting on each grain, isolate the contact interactions, and investigate how forces are distributed and transmitted in a barchan dune. We present force maps and probability density functions (PDFs) for values in the streamwise and spanwise directions, and show that stronger forces are experienced by grains at neither the crest nor leading edge of the barchan, but in positions just upstream the dune centroid on the periphery of the dune. We show also that a great part of grains undergo longitudinal forces of the order of 10$^{-7}$ N, with negative values around the crest, resulting in decelerations and grain deposition in that region. These data show that the force distribution tends to route a great part of grains toward the crest and horns of subaqueous barchans, being fundamental to comprehend their morphodynamics. However, to the best of the authors' knowledge, they are not accessible from current experiments, making of our results an important step toward understanding the behavior of barchan dunes.
\end{abstract}

\pacs{}

\maketitle 

\section{INTRODUCTION}
\label{sec:intro}

Barchan dunes, or simply barchans, are crescent-shaped dunes resulting from the transport of grains, usually sand, by a one-directional fluid flow in a mode of transport called bedload, in which grains roll, slide or effectuate small jumps maintaining contact with a fixed part of the bed \cite{Bagnold_1, Andreotti_1, Charru_5, Courrech}. For an initial granular pile, grains are eroded on regions upstream the pile crest and settle at the crest. The fluid flow then separates at the crest region and a recirculation bubble that forms just downstream the crest shapes a lee face with a curved base and horns pointing downstream \cite{Herrmann_Sauermann, Hersen_3, Alvarez, Alvarez3}. Under one-directional flow and bedload conditions, barchan dunes are robust bedforms that can be found in different environments and scales \cite{Hersen_1, Claudin_Andreotti}. For example, they are found in oil and water pipelines, rivers, Earth's deserts, and other planetary environments, their scales varying from the decimeter and minute under water \cite{Franklin_2, Franklin_8, Alvarez} to the kilometer and millennium on Mars \cite{Claudin_Andreotti, Parteli2}, passing by the hundred of meters and year on Earth's deserts \cite{Bagnold_1, Hersen_1}.

The large time scales of aeolian and martian barchans compared to the aquatic case make of subaqueous barchans the ideal object of study. Therefore, experiments have been conducted in water tanks and channels, where measurements were made at both the dune \cite{Hersen_1, Endo, Franklin_8, Hori, Reffet, Alvarez} and grain scales \cite{Alvarez3, Wenzel, Alvarez4}, and from which length and time scales of barchans and typical trajectories and velocities of moving grains were obtained. In particular, Wenzel and Franklin \cite{Wenzel} and Alvarez and Franklin \cite{Alvarez3, Alvarez4} are the only experimental studies showing the distribution of grain velocities over the barchan surface and trajectories of grains migrating to different parts of the barchan dune. For the forces on individual grains within a barchan dune, there are currently no experimental results, the acquisition of these data being highly difficult.

In addition to experiments in water, numerical simulations have been employed for the study of dunes. Also for numerical investigations, in particular those at the grain scale, subaqueous barchans are interesting since the number of involved grains is much smaller than in other environments. The first numerical investigations were on aeolian dunes and made use, initially, of continuum models for the grains \cite{Sauermann_4, Herrmann_Sauermann, Kroy_A, Kroy_C, Kroy_B, Schwammle, Parteli4}, and more recently of simplified discrete models such as the cellular automaton \cite{Narteau, Zhang_D}. The most recent investigations were focused on subaqueous bedload and bedforms and used Euler-Lagrange methods, such as computational fluid dynamics - discrete element method (CFD-DEM) \cite{Schmeeckle, Kidanemariam, Kidanemariam2, Kidanemariam3, Liu, Sun, Pahtz_3, Pahtz_4}. In particular, Kidanemariam and Uhlmann \cite{Kidanemariam2, Kidanemariam, Kidanemariam3} used direct numerical simulations (DNS) for the fluid and DEM for the grains, both coupled by immersed boundary (IB), which is, currently, the most accurate technique, fully solving the flow around each grain. However, while it captures all turbulence scales down to Kolmogorov scale, the computational cost is exceedingly high and the time required for obtaining developed barchans is seldom reached \cite{Colombini}.

In a recent paper \cite{Alvarez5}, we presented numerical computations of the growth and evolution of subaqueous barchans carried out at the grain scale, where we coupled DEM with large eddy simulation (LES). LES, although less accurate than DNS and needing turbulence models, is able to compute the flow around dunes at a much lower computational cost. The simulations captured well the evolution of an initial pile toward a barchan dune in both the bedform and grain scales, with the same characteristic time and lengths observed in previous experiments \cite{Alvarez, Alvarez3, Alvarez4}. In addition to reproducing accurately previous experimental data, the numerical results revealed in detail quantities not accessible from experiments, such as the resultant force acting on each grain. However, although such quantities are important to understand the behavior of barchans, an analysis of the forces experienced by each grain within a barchan dune is still missing.

In this paper, we investigate how forces on grains are transmitted within a subaqueous barchan. Based on the CFD-DEM computations of Alvarez and Franklin \cite{Alvarez5}, we plot maps showing the distribution of grain forces in an isolated barchan and compute probability density functions (PDFs) for values in the streamwise and spanwise directions. We show that force distributions tend to route a great part of grains toward the crest and horns of subaqueous barchans, with the stronger forces experienced by grains occurring at neither the crest nor leading edge of the barchan, but in positions just upstream the dune centroid on the periphery of the dune. We show also that a great part of grains undergo forces in the streamwise direction of the order of 10$^{-7}$ N, and that on the crest they have negative values, resulting in decelerations and grain deposition in the crest region. Finally, we found that around 9\% of grains migrate to each horn and 13\% to the dune crest. Our results show, for the first time, the values of the resultant force acting on each particle for all grains within a barchan dune. To the authors' knowledge, these data are not accessible from current experiments and represent a new step for understanding the motion of grains over the dune surface as well as the load experienced by grains below the bedload layer. The present data are thus fundamental to comprehend the morphodynamics of barchans, being linked to the distribution of grains within a barchan dune.

\section{METHODS}
\label{sec:methods}

Our simulations coupled DEM with LES to compute numerically the growth and evolution of single barchans from a conical pile. For that, we used the open-source code \mbox{CFDEM} \cite{Goniva} (www.cfdem.com), which couples the open-source CFD code OpenFOAM with the open-source DEM code LIGGGHTS \cite{Kloss, Berger}. We describe briefly in the following the used model and numerical setup, a detailed description being found in Alvarez and Franklin \cite{Alvarez5}.

The DEM part computes the dynamics of solid particles in a Lagrangian framework by using the linear and angular momentum equations, Eqs. \ref{Fp} and \ref{Tp}, respectively,

\begin{equation}
m_{p}\frac{d\vec{u}_{p}}{dt}= \vec{F}_{p}\, ,
\label{Fp}
\end{equation}

\begin{equation}
I_{p}\frac{d\vec{\omega}_{p}}{dt}=\vec{T}_{c}\, ,
\label{Tp}
\end{equation}

\noindent where, for each grain, $m_{p}$ is the mass, $\vec{u}_{p}$ is the velocity, $I_{p}$ is the moment of inertia, $\vec{\omega}_{p}$ is the angular velocity, $\vec{T}_{c}$ is the resultant of contact torques between solids, and $\vec{F}_{p}$ is the resultant force,

\begin{equation}
\vec{F}_{p} = \vec{F}_{fp} + \vec{F}_{c} + m_{p}\vec{g} \, ,
\label{Fp2}
\end{equation}

\noindent  where $\vec{F}_{fp}$ is the resultant of fluid forces acting on a grain, $\vec{F}_{c}$ is the resultant of contact forces between solids, and $\vec{g}$ is the acceleration of gravity. In our simulations, we consider that $\vec{F}_{fp}$ is the sum of components given by the fluid drag, fluid stresses and added mass, and we neglect the Basset, Saffman and Magnus forces once they are usually considered of lesser importance in CFD-DEM simulations \cite{Zhou}. For the angular momentum, Eq. \ref{Tp}, we neglect torques caused by the direct action of the fluid since the term due to contacts is usually much higher \cite{Tsuji, Tsuji2, Liu}.

The CFD part computes the dynamics of the fluid phase in an Eulerian framework, by solving the incompressible mass and momentum equations, Eqs. \ref{mass} and \ref{mom}, respectively,

\begin{equation}
\nabla\cdot\vec{u}_{f}=0 \, ,
\label{mass}
\end{equation}

\begin{equation}
\frac{\partial{\rho_{f}\vec{u}_{f}}}{\partial{t}} + \nabla \cdot (\rho_{f}\vec{u}_{f}\vec{u}_{f}) = -\nabla P + \nabla\cdot \vec{\vec{\tau}} + \rho_{f}\vec{g} - \vec{f}_{fp} \, ,
\label{mom}
\end{equation}

\noindent where $\vec{u}_{f}$ is the fluid velocity, $\rho_{f}$ is the fluid density, and $\vec{f}_{fp}$ is the resultant of fluid forces acting on each grain, $\vec{F}_{fp}$, by unit of fluid volume.

For the DEM, we considered a Hertzian model for which we used the parameters listed in Tab. \ref{tabsim}. The boundary conditions for the grains were solid walls at the top and bottom walls, no mass entering at the inlet, and free exit at the outlet. As initial condition, grains were poured from above, falling freely in still water until they settled completely.

\begin{table}[ht]
	\begin{center}
	\caption{Parameters used in the DEM computations}
	\begin{tabular}{c c}
		\hline\hline
		Initial number of particles & 4 $\times$ 10$^4$\\
		Particle diameter $d$ (mm)  & 0.5 \\
		Particle density $\rho_p$ (kg/m$^3$) & 2500 \\
		Restitution coefficient $e$ & 0.1 \\
		Friction coefficient $\mu_{fr}$ & 0.6\\
		Young's Modulus $E$ (MPa) & 5 \\
		Poisson ratio $\sigma$ & 0.45 \\
		Time step (s) & 5 $\times$ $10^{-6}$\\
		\hline
	\end{tabular}
		\label{tabsim}
    \end{center}
\end{table}

For the CFD, we used LES with the wall-adapting local eddy-viscosity (WALE) model \cite{Nicoud}, and the domain was set to 0.3 $\times$ 0.05 $\times$ 0.16 m in the streamwise, $x$, wall-normal, $y$, and spanwise, $z$, directions, respectively. The $x$ and $z$ directions were divided into 150 and 160 segments that were uniform in size, whereas the $y$ direction was divided into 150 unevenly spaced segments. Our simulations were performed for two different flow conditions, corresponding to channel Reynolds numbers based on the cross-sectional mean velocity $U$ and channel height $2\delta$, Re = $U 2\delta \nu^{-1}$, of 1.47$\times$10$^4$ and 1.82$\times$10$^4$, and to Reynolds numbers based on the shear velocity $u_*$, Re$_*$ = $u_* \delta \nu^{-1}$, of 420 and 506, where $\nu$ is the kinematic viscosity of the fluid. The resulting Shields numbers, $\theta$ = $u_*^2/ \left( \left( \rho_p \rho_{f} ^{-1} - 1 \right) g d \right)$, were of 0.04 and 0.06, where $\rho_p$ and $d$ are the density and diameter of solid particles. For the two flow conditions, the  grid spacings in the streamwise and spanwise directions scaled in inner wall units ($\nu u_*^{-1}$) were of 33.6 and 40.4, and 16.8 and 20.2, respectively, and the normalized grid spacings in the wall-normal direction at the first point were of 0.91 and 1.10. The boundary conditions for the fluid were impermeability and no-slip conditions at the top and bottom walls, and periodic conditions in the longitudinal and transverse directions. The initial condition was based on single phase flows computed prior to simulations with grains, from which the final realization was used as the initial condition.

From the simulations, we obtained the forces on each grain, and we investigate now how forces are distributed within the barchan. A layout of the numerical setup is available in the supplementary material and numerical data (in terms of forces) from our simulations are available in Mendeley Data \cite{Supplemental3}.

\section{RESULTS}
\label{sec:results}

We present next values of forces acting on each grain within a barchan dune. Because data for forces experienced by individual grains within a barchan dune are still missing, we cannot directly compare our results with previous works in terms of forces. Instead, after presenting the values of forces, we analyze our results in terms of trajectories, obtained experimentally by Alvarez and Franklin \cite{Alvarez3, Alvarez4}.

 \begin{figure*}
 	\begin{minipage}{0.49\linewidth}
 		\begin{tabular}{c}
 			\includegraphics[width=0.90\linewidth]{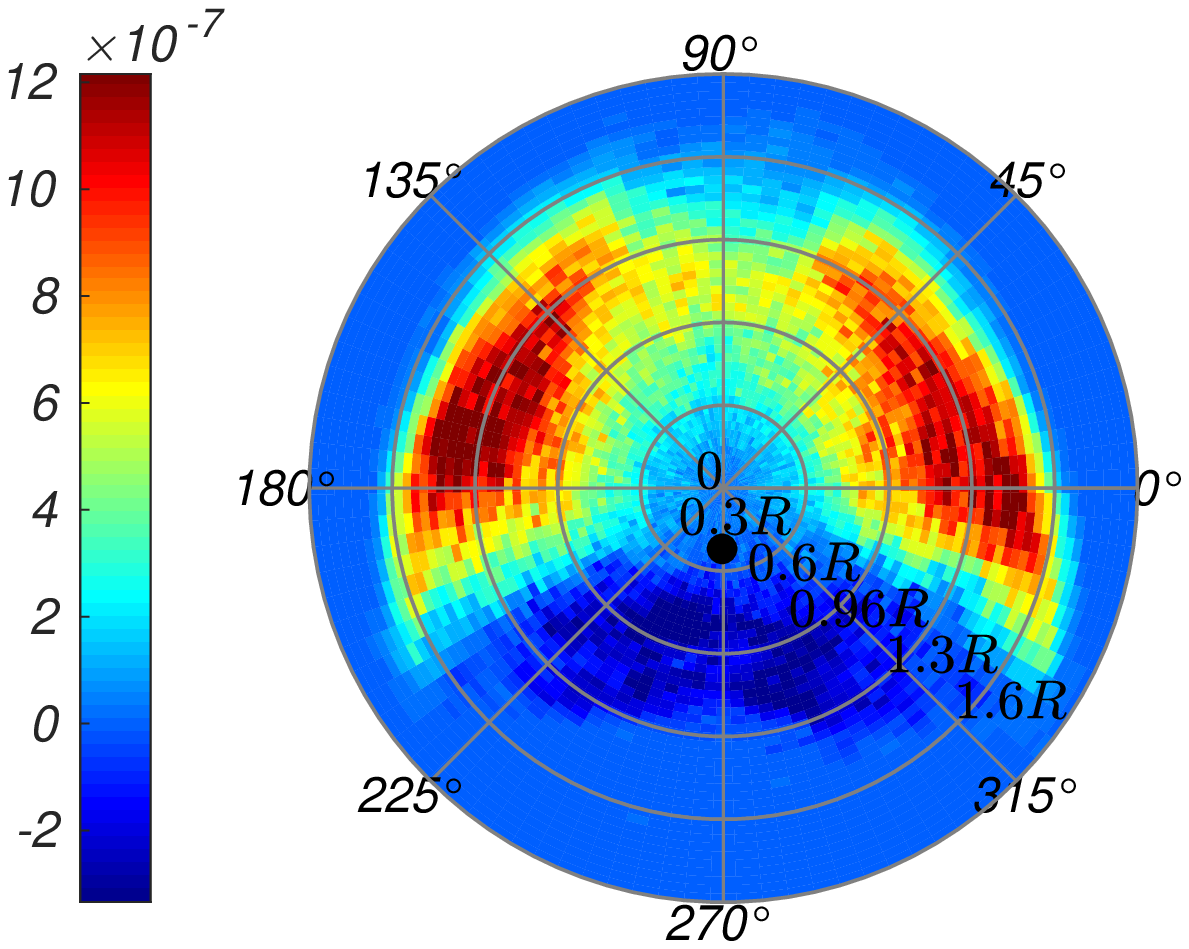}\\
 			(a)
 		\end{tabular}
 	\end{minipage}
 	\hfill
 	\begin{minipage}{0.49\linewidth}
 		\begin{tabular}{c}
 			\includegraphics[width=0.90\linewidth]{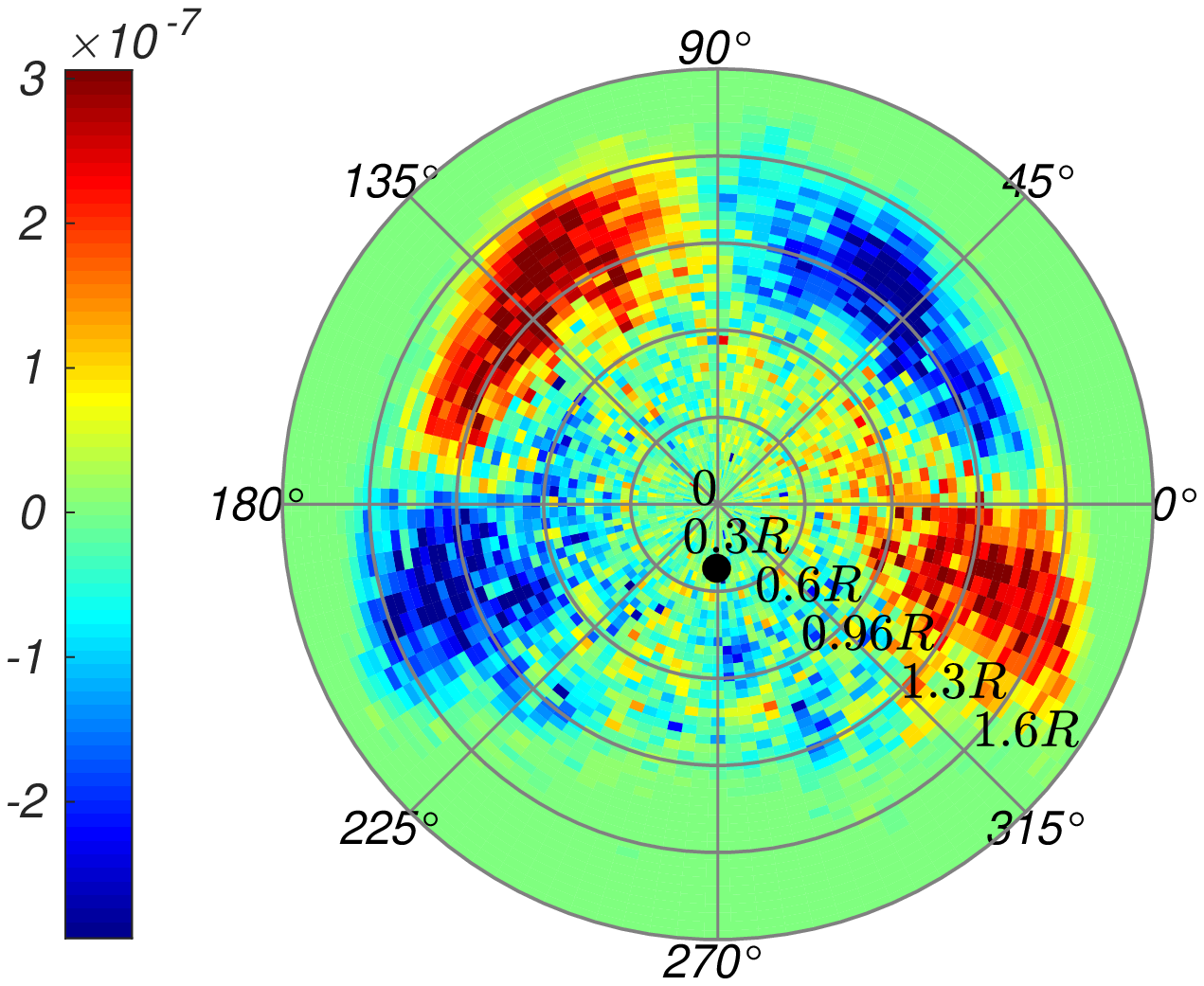}\\
 			(b)
 		\end{tabular}
 	\end{minipage}
	\hfill
	\caption{Maps of the resultant force on grains composing an evolving barchan dune ($t<2.5t_c$) in the (a) streamwise and (b) spanwise directions, $F_{px}$ and $F_{pz}$, respectively. This figure corresponds to $Re$ = 1.47 $\times$ 10$^4$, and the solid circle in the maps indicates the crest position. Values in the colorbar are in N.}
	\label{fig:map_force}
 \end{figure*}

 \begin{figure*}
 	\begin{minipage}{0.49\linewidth}
 		\begin{tabular}{c}
 			\includegraphics[width=0.90\linewidth]{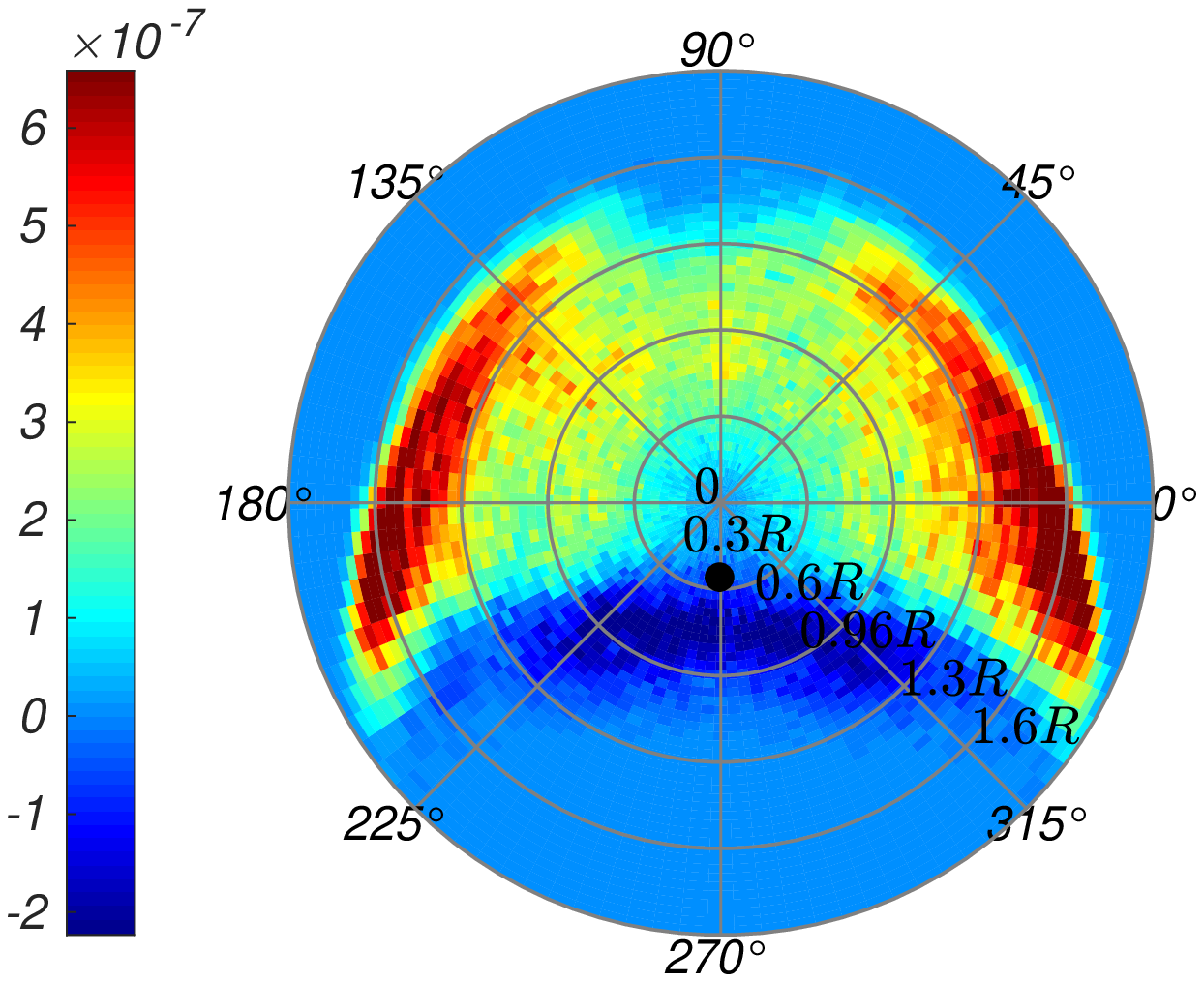}\\
 			(a)
 		\end{tabular}
 	\end{minipage}
 	\hfill
 	\begin{minipage}{0.49\linewidth}
 		\begin{tabular}{c}
 			\includegraphics[width=0.90\linewidth]{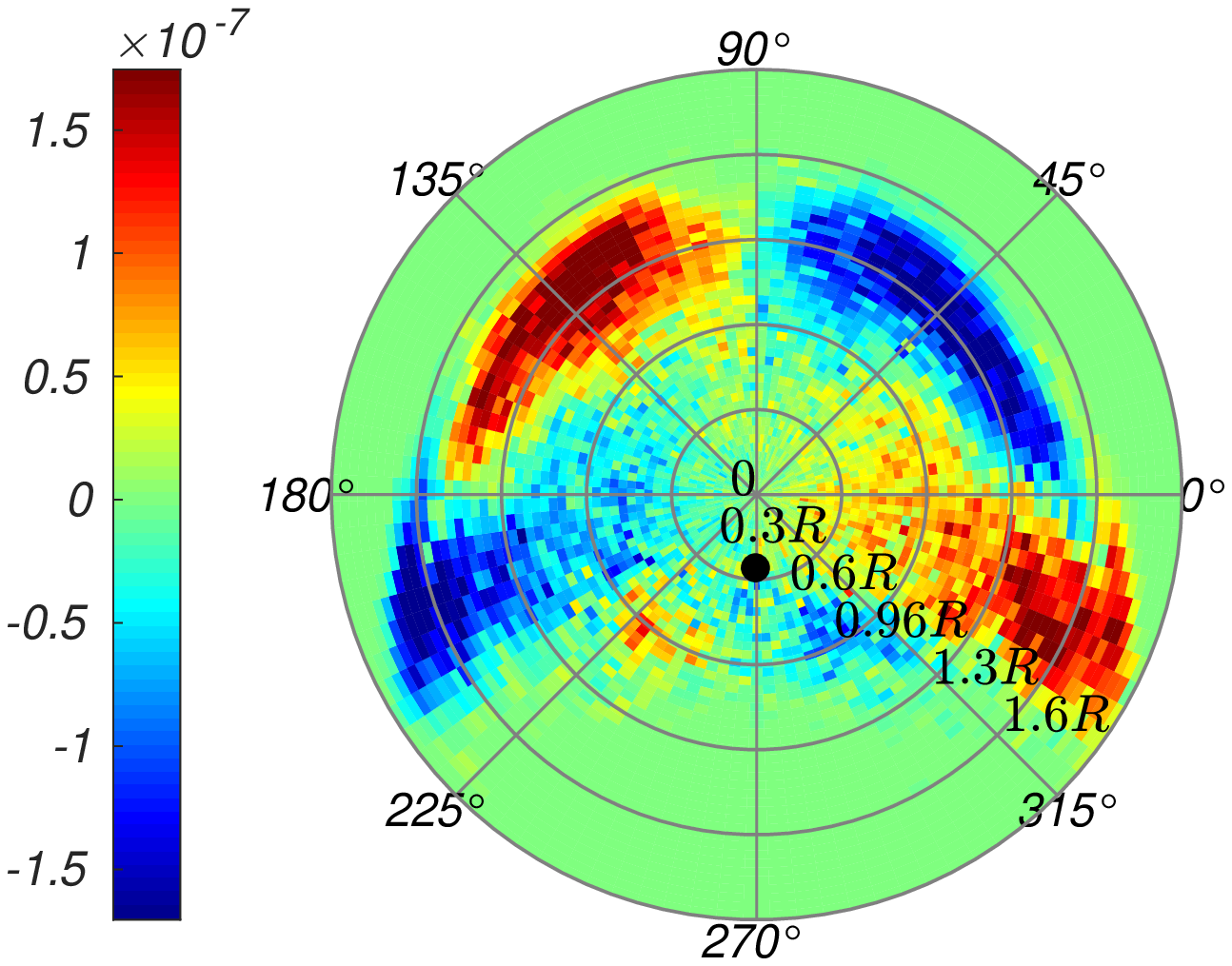}\\
 			(b)
 		\end{tabular}
 	\end{minipage}
	\hfill
	\caption{Maps of the resultant force on grains composing a developed barchan dune ($t>2.5t_c$) in the (a) streamwise and (b) spanwise directions, $F_{px}$ and $F_{pz}$, respectively. This figure corresponds to $Re$ = 1.47 $\times$ 10$^4$, and the solid circle in the maps indicates the crest position. Values in the colorbar are in N.}
	\label{fig:map_force_longer}
 \end{figure*}

Figures \ref{fig:map_force} and \ref{fig:map_force_longer} show maps of the resultant force on grains composing a barchan dune, $\vec{F}_{p}$, in both the streamwise and spanwise directions for growing ($t<2.5t_c$) and developed ($t>2.5t_c$) barchans, respectively, where $t_c$ is a characteristic time for the displacement of barchans computed as the length of the bedform divided by its celerity \cite{Alvarez}. Figures \ref{fig:map_force}(a) and \ref{fig:map_force_longer}(a) correspond to the resultant force in the streamwise direction, $F_{px}$, and Figs. \ref{fig:map_force}(b) and \ref{fig:map_force_longer}(b) in the spanwise direction, $F_{pz}$. The maps of Figs. \ref{fig:map_force} and \ref{fig:map_force_longer} are in polar coordinates with origin at the dune centroid, the water flow direction is 270$^{\circ}$, and $R$ is the radius of the initial conical pile \cite{Alvarez3, Alvarez4}. The mean forces were averaged over all grains, including those inside the barchan, and from the $t=0$ (initial condition) until $t=2.5t_c$ for Fig. \ref{fig:map_force} and from $t=2.5t_c$ to $t=5.0t_c$ for Fig. \ref{fig:map_force_longer}. For $F_{px}$, positive values point downstream and negative upstream, while for $F_{pz}$ positive values point to the left (180$^{\circ}$) and negative to the right (0$^{\circ}$), and they were computed for $Re$ = 1.47 $\times$ 10$^4$.

From Figs. \ref{fig:map_force}(a) and \ref{fig:map_force_longer}(a), we observe that forces on grains point upstream at the crest, indicating that grains eroded upstream and migrating toward the crest decelerate and settle there. In the region downstream the crest, in between evolving or developed horns (215$^{\circ}$ $\lesssim$ $\alpha$ $\lesssim$ 325$^{\circ}$, where $\alpha$ is the angle formed with 0$^{\circ}$), forces on grains point upstream, showing that grains arriving in that region are subjected to the recirculation bubble and stay trapped close to the lee face (forming or already formed). Figures \ref{fig:map_force}(a) and \ref{fig:map_force_longer}(a) show also that forces in the downstream direction are stronger neither at the leading edge of the bedform nor along the symmetry line in the streamwise direction, but toward the lateral flanks of the dune, in the regions within 120$^{\circ}$ $\lesssim$ $\alpha$ $\lesssim$ 210$^{\circ}$ and 60$^{\circ}$ $\lesssim$ $\alpha$ $\lesssim$ 330$^{\circ}$. Together, the maps for both $F_{px}$ and $F_{pz}$ of evolving and developed barchans show that, for distances farther than 0.6$R$ from the centroid, grains within 10$^{\circ}$ $\lesssim$ $\alpha$ $\lesssim$ 80$^{\circ}$ and 100$^{\circ}$ $\lesssim$ $\alpha$ $\lesssim$ 170$^{\circ}$ are forced outwards (with respect to the streamwise centerline) and downstream, while those within 0$^{\circ}$ $\lesssim$ $\alpha$ $\lesssim$ 330$^{\circ}$ and 210$^{\circ}$ $\lesssim$ $\alpha$ $<$ 180$^{\circ}$ are forced inwards and downstream. This corroborates our previous experimental findings \cite{Alvarez3, Alvarez4} that in the subaqueous case barchan horns are formed and sustained with grains coming from upstream regions on the periphery of the bedform. Although previous works on morphodynamics considered the fluid flow and inertial effects to explain the growth of bedforms based on the flux of grains \cite{Engelund_Fredsoe, Charru_5}, the resultant forces on individual grains, responsible for their trajectories, had never been shown before.

We observe forces that tend to entrain grains to the crest and horns of both evolving and developed barchans, which corroborates the observations made in experiments on subaqueous barchans \cite{Alvarez3, Alvarez4} (see the supplementary material for a movie from our numerical simulations showing the instantaneous values of the resultant force on each grain). We estimated the granular flux by computing the number of grains crossing a barchan cross section ($x$ plane) at $x=0$ (crest position) for a duration equivalent to 5.0$t_c$. We then counted the number of grains going to the crest and each of the horns, and computed the respective percentages. We found that around 9\% of grains migrate to each horn (so that 18\% of grains migrate to horns) and 13\% to the crest, the remainder grains migrating to either the regions between the crest and one of the horns, where they settle before falling by avalanches, or going around the dune until reaching the horn tips, from where they are entrained downstream.

 \begin{figure*}
 	\begin{minipage}{0.49\linewidth}
 		\begin{tabular}{c}
 			\includegraphics[width=0.90\linewidth]{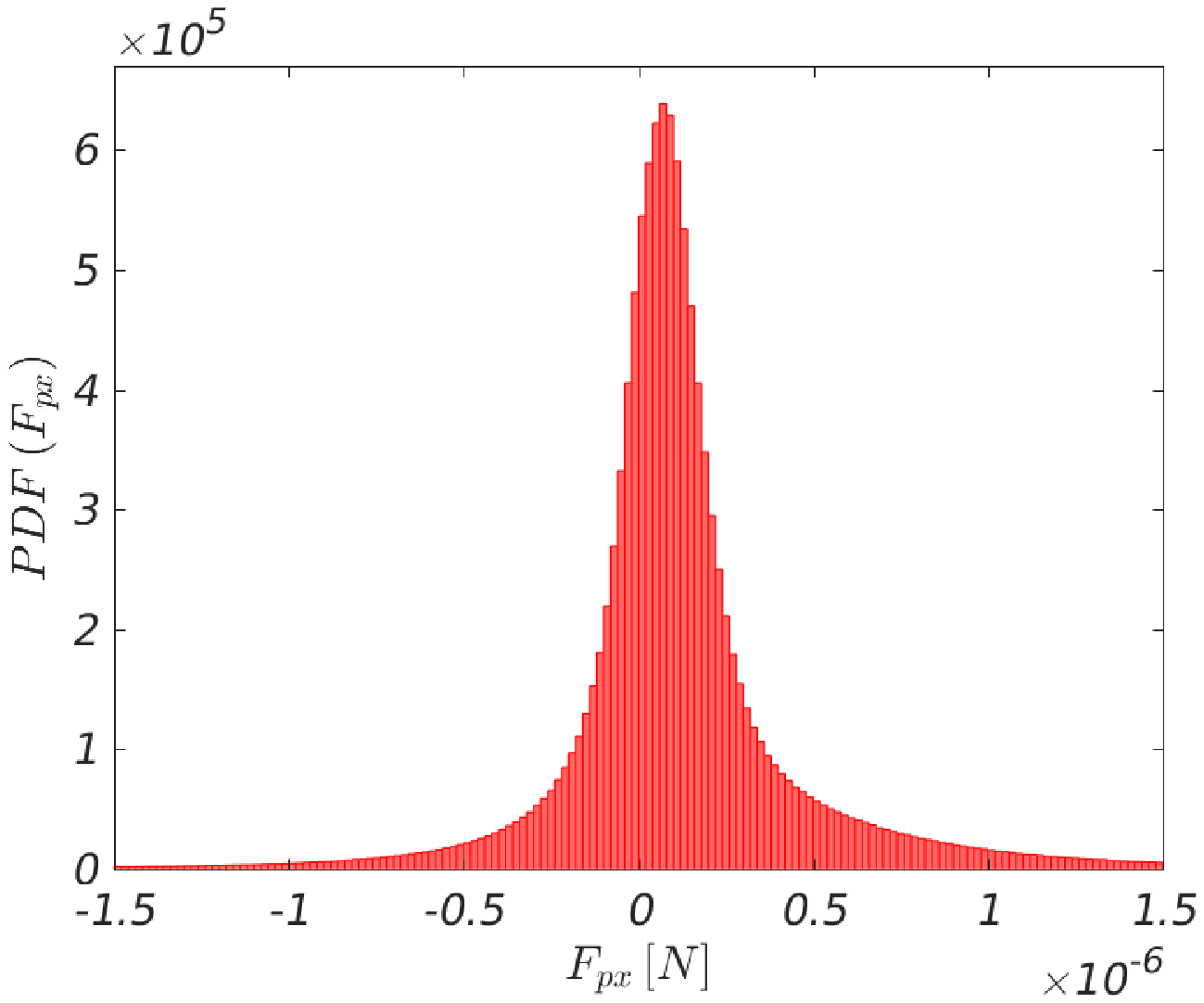}\\
 			(a)
 		\end{tabular}
 	\end{minipage}
 	\hfill
 	\begin{minipage}{0.49\linewidth}
 		\begin{tabular}{c}
 			\includegraphics[width=0.90\linewidth]{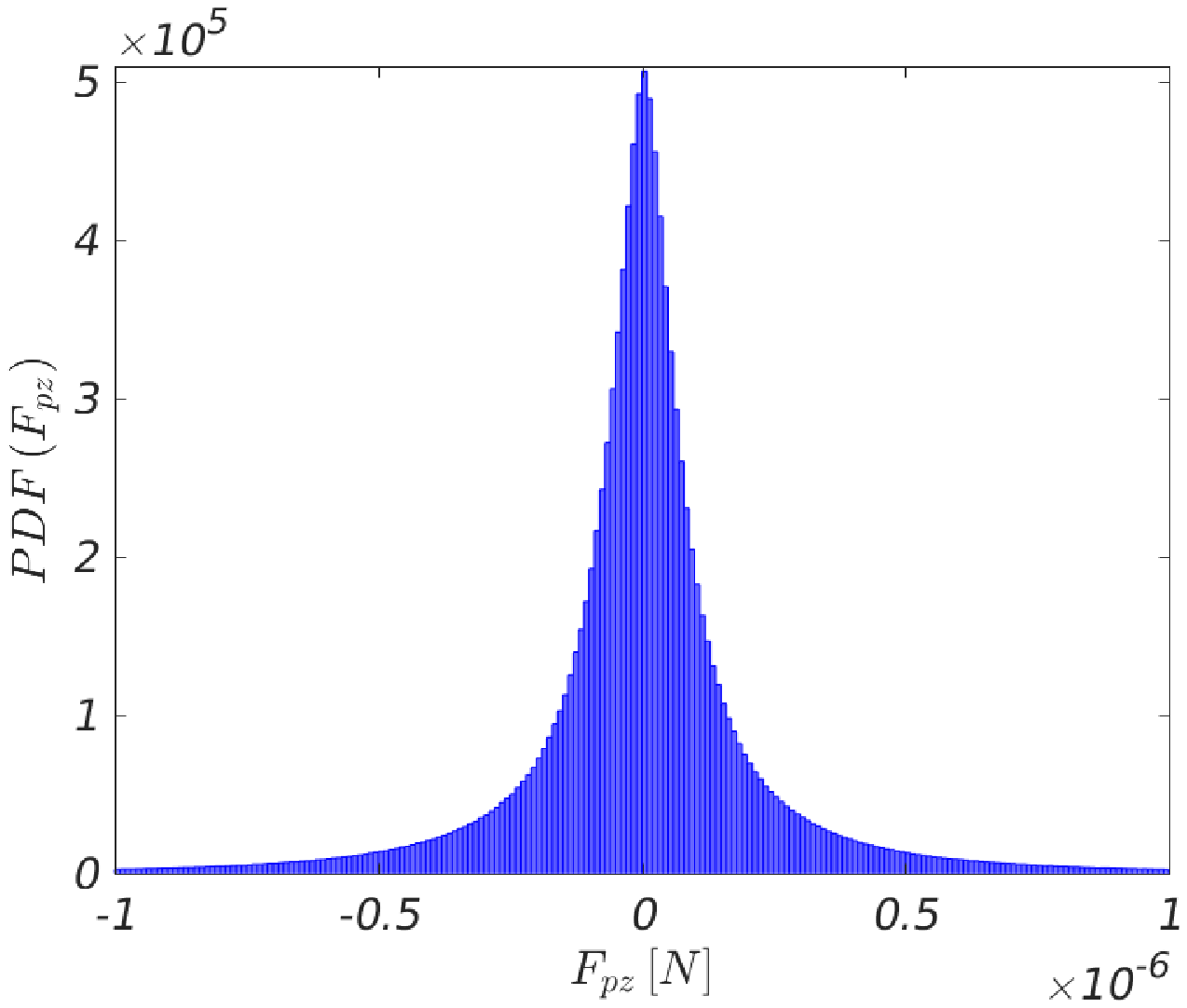}\\
 			(b)
 		\end{tabular}
 	\end{minipage}
	\hfill
	\caption{PDFs of the resultant forces in the (a) streamwise and (b) spanwise directions. $Re$ = 1.47 $\times$ 10$^4$.}
	\label{fig:pdf_total_force}
 \end{figure*}

 \begin{figure*}
 	\begin{minipage}{0.49\linewidth}
 		\begin{tabular}{c}
 			\includegraphics[width=0.90\linewidth]{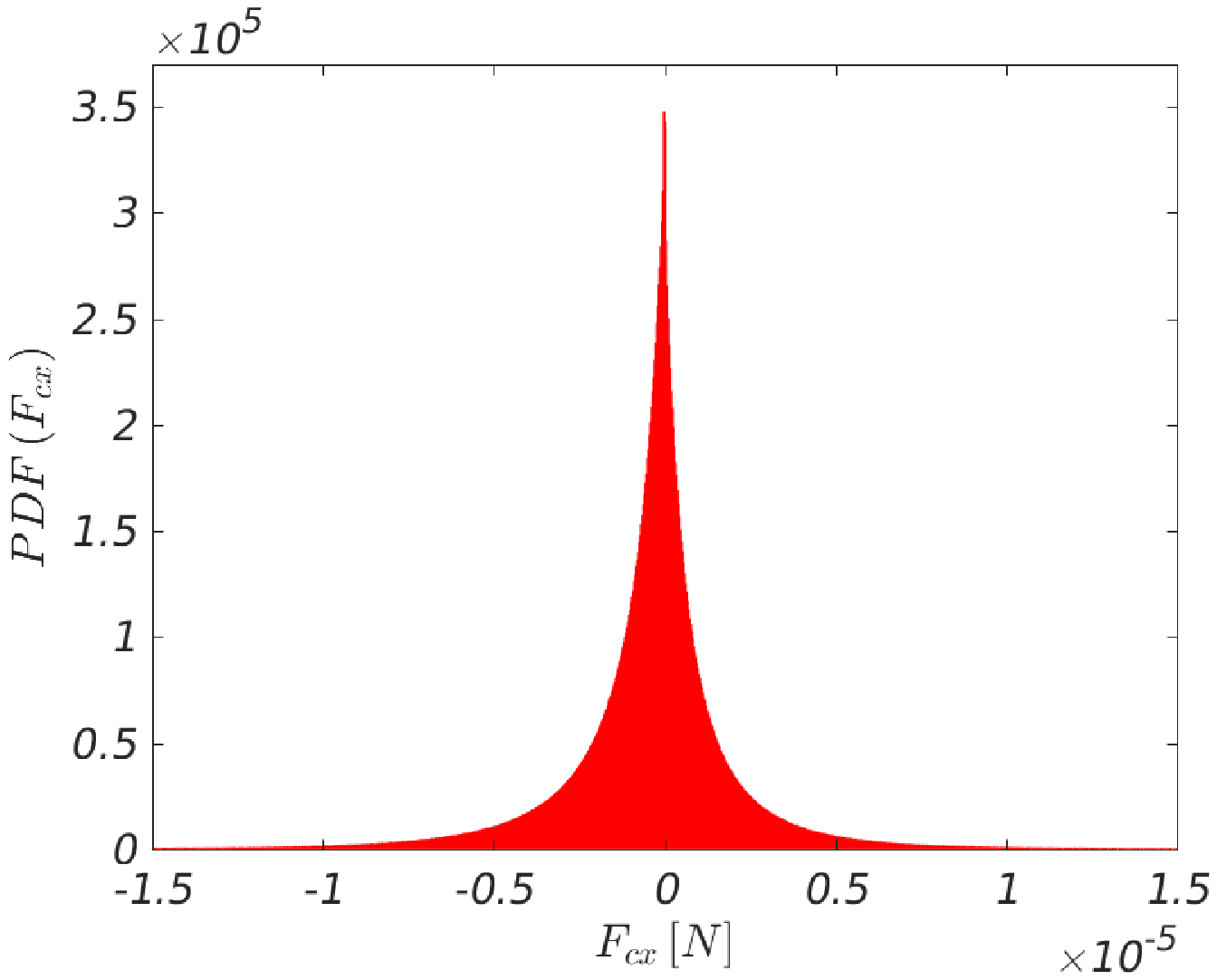}\\
 			(a)
 		\end{tabular}
 	\end{minipage}
 	\hfill
 	\begin{minipage}{0.49\linewidth}
 		\begin{tabular}{c}
 			\includegraphics[width=0.90\linewidth]{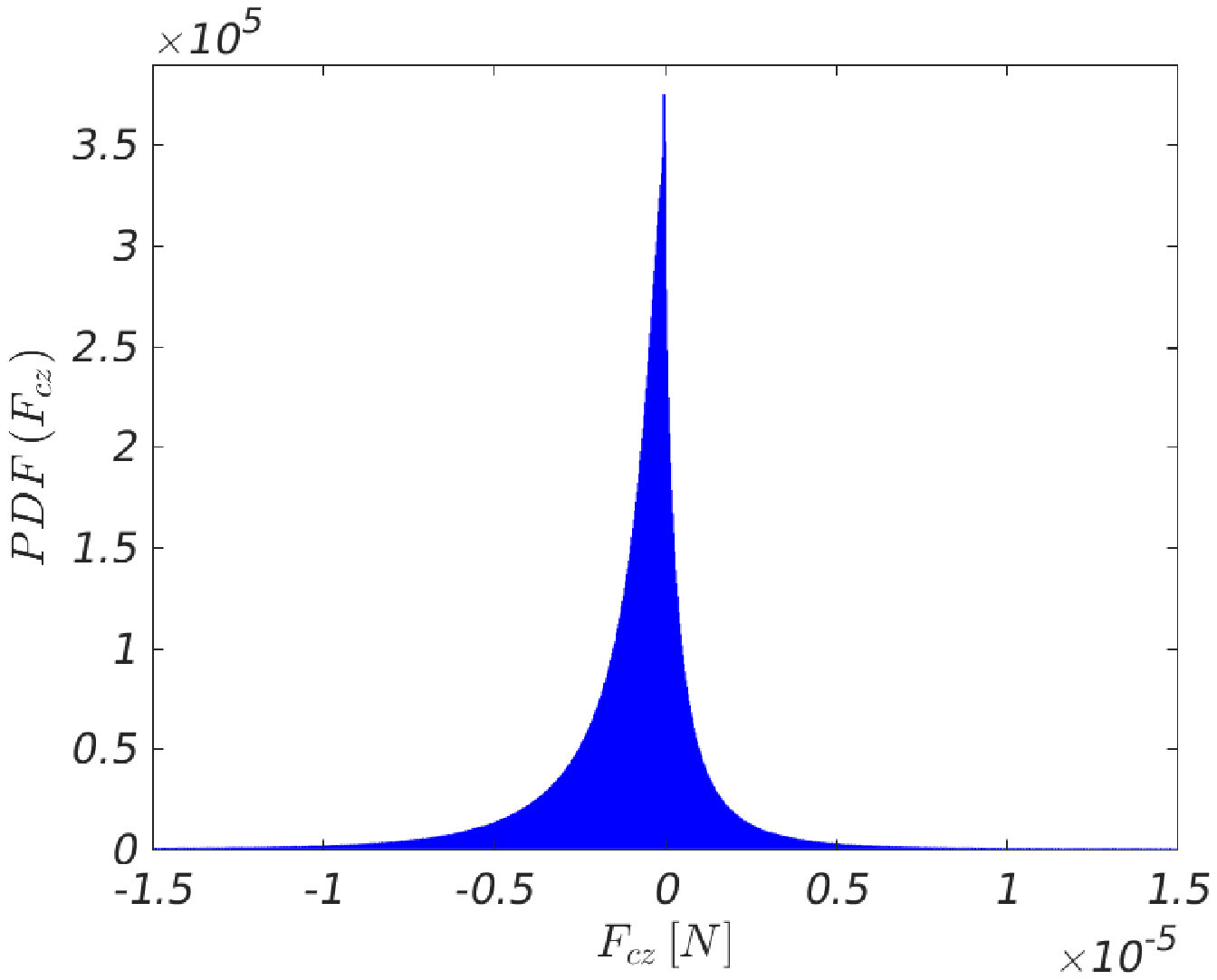}\\
 			(b)
 		\end{tabular}
 	\end{minipage}
	\hfill
	\caption{PDFs of contact forces in the (a) streamwise and (b) spanwise directions. $Re$ = 1.47 $\times$ 10$^4$.}
	\label{fig:pdf_force}
 \end{figure*}

We computed, for all grains, the probability density functions (PDFs) of the resultant force from $t$=0 to $t$=5.0$t_c$, presented in Figs. \ref{fig:pdf_total_force}(a) and \ref{fig:pdf_total_force}(b) for values in the streamwise and spanwise directions, respectively. We observe that the PDF of $F_{pz}$ is symmetric, peaked at zero and with a width approximately equal to that of $F_{px}$, indicating that in average grains have considerable spanwise forces, but with a zero mean (and also most probable value). This corroborates the large spanwise displacements of grains observed in subaqueous barchans \cite{Alvarez3, Alvarez4}, following circular trajectories and migrating toward the horns (symmetrically with respect to the streamwise centerline). For $F_{px}$, the PDF is not symmetric and is peaked at a value of the order of 10$^{-7}$ N, which is the same order of magnitude of the mean value. In addition, Fig. \ref{fig:pdf_total_force}(a) shows a probability of 86.3\% of finding $F_{px}$ within -0.5 $\times$ 10$^{-6}$ and 0.5 $\times$ 10$^{-6}$ N, and that 5.5\% of grains undergo values of $F_{px}$ higher than their relative weight (of approximately 1 $\times$ 10$^{-6}$ N), corresponding to highly accelerated and decelerated grains. In order to investigate the transmission of forces by contacts, we separated the contact from the other forces (Eq. \ref{Fp2}) for all grains within the bedform and computed the PDFs for values in the streamwise ($F_{cx}$) and spanwise ($F_{cz}$) directions, presented in Figs. \ref{fig:pdf_force}(a) and \ref{fig:pdf_force}(b), respectively. Both PDFs are peaked at zero, as expected since the action of one grain is the reaction on others, with most of values ranging from -0.5 $\times$ 10$^{-5}$ to 0.5 $\times$ 10$^{-5}$N, such extremes being of the order of the grain's weight. In addition, we observe that the PDF in the spanwise direction has a width comparable to that in the streamwise direction. Because very few grains experience resultant forces in the streamwise direction stronger than their weight, whereas all of them undergo gravity, most of values transmitted by contacts and presented in Figs. \ref{fig:pdf_force}(a) and \ref{fig:pdf_force}(b) are due to gravity. However, while $F_{cx}$ presents a small asymmetry, which could be inferred to contact forces with the wall being neglected in the PDF, $F_{cz}$ has a relatively larger asymmetry. We believe that these asymmetries are caused by irregularities in the contact network and the fact that Figs. \ref{fig:pdf_force}(a) and \ref{fig:pdf_force}(b) are based upon one particular simulation. Networks of contact forces corroborate that supposition, and are available in the supplementary material.

\begin{figure}
\begin{center}
		\includegraphics[width=0.6\columnwidth]{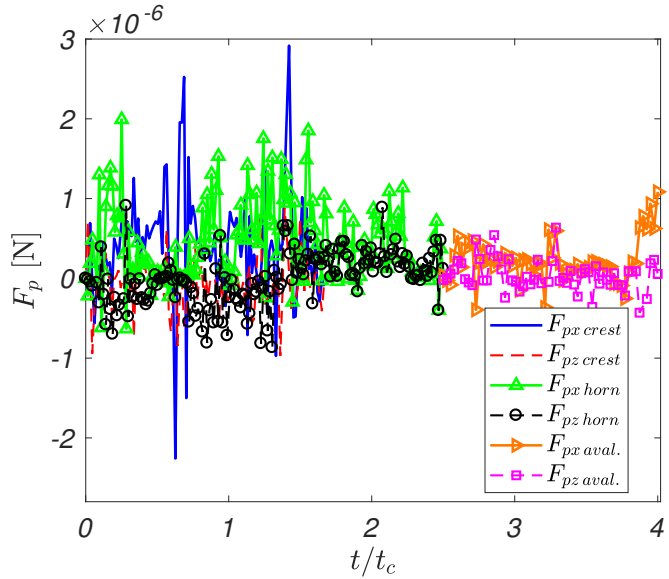}\\	
\end{center}
    \caption{Resultant force on some individual grains migrating to the crest, to one of the horns, and falling by avalanches, as functions of the normalized time. $Re$ = 1.47 $\times$ 10$^4$.}
    \label{fig:lagrangian}
\end{figure}

Finally, Fig. \ref{fig:lagrangian} shows examples of Lagrangian trackings of the resultant force acting on individual grains migrating to the crest, to one of the horns, and falling by avalanches. Figure \ref{fig:lagrangian} presents one example for each trajectory, but they represent well other grains following similar trajectories. For the grain migrating to the horn, it took 1.8$t_c$ to reach the horn, and was tracked for additional 0.7$t_c$ (2.5$t_c$ being the characteristic time for growth of subaqueous barchans \cite{Alvarez}), while the grain migrating to the crest spent approximately 1.6$t_c$ to arrive there. The grain tracked during an avalanche started falling at 2.5$t_c$ and took 1.1$t_c$ to arrive at the dune base, after which it was buried. In particular, we note large variations in $F_{px}$ for grains migrating to horns and crest, with positive average values. These variations express the intermittent motion of grains in subaqueous bedload \cite{Lajeunesse, Penteado}. For grains involved in avalanches, fluctuations are smaller and due mainly to contacts during their fall.

Our results consist in new information, not accessible from previous experiments or simulations, on the resultant force acting on each grain, and, therefore, on how forces are distributed within a subaqueous barchan dune. However, differences with respect to aeolian and martian dunes are expected since grains within a bedload layer move differently depending on the fluid state. When the fluid is a liquid, grains move by rolling and sliding and follow closely the fluid flow, while for gases grains move by saltation and reptation, those in saltation following ballistic flights in the main flow direction. In spite of these differences, our results represent an important step toward understanding the morphodyamics of barchans.

\section{CONCLUSIONS}
\label{sec:conclusions}

Based on CFD-DEM computations, we measured the resultant force acting on each grain for all grains composing a barchan dune. We obtained maps of force distributions within evolving and developed barchans, PDFs of the magnitude of the resultant and contact forces, and the time evolution of forces on tracked grains. We confirmed that the resultant force on grains acts in their entrainment toward the crest and horns of subaqueous barchans. In particular, we showed that stronger forces on grains occur at neither the crest nor leading edge of the barchan, but in positions just upstream the dune centroid on the periphery of the dune, which corroborates the trajectories of grains migrating to horns reported in the literature \cite{Alvarez3, Alvarez4}. We showed also that a great part of grains undergo longitudinal forces of the order of 10$^{-7}$ N (around 86\% of grains experience resultant forces within -0.5 $\times$ 10$^{-6}$ and 0.5 $\times$ 10$^{-6}$ N), with negative values around the crest, resulting in decelerations and grain deposition in that region. Finally, we found that around 18\% of grains migrate to horns and 13\% to the crest, the remainder grains migrating to either the regions between the crest and one of the horns (where they settle before falling by avalanches) or going around the dune until reaching the horn tips (from where they are entrained downstream). The present results provide new insights into barchan morphology and how grains are distributed within the dune.

\section*{SUPPLEMENTARY MATERIAL}
See the supplementary material for a layout of the numerical setup, PDFs of the fluid forces on grains, networks of contact forces within the bedform, and a movie from our numerical simulations showing the instantaneous values of the resultant force on each grain.

\section*{DATA AVAILABILITY}
The data that support the findings of this study are openly available in Mendeley Data at http://dx.doi.org/10.17632/g666hxgrty.1.

\begin{acknowledgments}
Carlos A. Alvarez is grateful to SENESCYT (Grant No. 2013-AR2Q2850) and to CNPq (Grant No. 140773/2016-9). Erick Franklin is grateful to FAPESP (Grant No. 2018/14981-7) and to CNPq (Grant No. 400284/2016-2) for the financial support provided. Carlos A. Alvarez is also grateful to Western University and the Emerging Leaders of the Americas Program, Global Affairs Canada. The authors would like to thank Prof. J. M. Floryan from Western University, Canada, for helpful discussions. SHARCNET (www.sharcnet.ca) provided part of computational resources used in the project.
\end{acknowledgments}

\bibliography{references}

\end{document}